\documentclass[aps,pre,10pt,onecolumn,notitlepage,superscriptaddress]{revtex4-1}

\usepackage{epsfig}
\usepackage{amssymb,amsmath}
\usepackage{array}

\usepackage{graphicx}
\usepackage{dcolumn}
\usepackage{bm}
\usepackage{lineno}
\usepackage{color}

\newcommand{\vect}[1] {{\bf #1}}

\newcommand{\beq}{\begin{equation}}
\newcommand{\eeq}{\end{equation}}
\newcommand{\beqa} { \begin{align} }
\newcommand{\eeqa} {\end{align}}
\newcommand{\beqs} {\begin{displaymath}}
\newcommand{\eeqs} {\end{displaymath}}
\newcommand{\beqas} {\begin{eqnarray*}}
\newcommand{\eeqas} {\end{eqnarray*}}
\newcommand{\et}{\emph{et al. }}
\newcommand{\pp}{\emph{P. phosphoreum}}
\newcommand{\be}{\begin{eqnarray}}
\newcommand{\ee}{\end{eqnarray}}
\newcommand{\bse}{\begin{subequations}}
\newcommand{\ese}{\end{subequations}}
\newcommand{\ba}[1]{\begin{align}#1\end{align}}
\newcommand{\bnum}{\begin{enumerate}}
\newcommand{\enum}{\end{enumerate}}
\newcommand{\bit}{\begin{itemize}}
\newcommand{\eit}{\end{itemize}}
\newcommand{\bc}{\begin{cases}}
\newcommand{\ec}{\end{cases}}

\newcommand{\gr}{\rho}

\newcommand{\f}{\frac}

\newcommand{\lap}{\triangle}
\newcommand{\nab}{\nabla}
\newcommand{\parder}[2]{\frac{\partial #1}{\partial #2}}
\newcommand{\parderd}[2]{\frac{\partial^2 #1}{\partial #2^2}}

\begin{document}

\title{Evaporation-driven convective flows in suspensions of non-motile bacteria}


\author{Jocelyn Dunstan}
\altaffiliation[Present address: ]{Public Health School, University of Chile, Santiago, Chile}
\affiliation{Department of Applied Mathematics and Theoretical Physics, Centre for Mathematical Sciences, University of Cambridge, 
Cambridge CB3 0WA, United Kingdom}
\author{Kyoung J. Lee}
\affiliation{Department of Physics, Korea University, Seoul, Korea}
\author{Simon F. Park}
\affiliation{Faculty of Health \& Medical Sciences, University of Surrey, Guildford GU2 5XH, Surrey, United Kingdom}
\author{Yongyun Hwang}
\affiliation{Department of Aeronautics, Imperial College, London, United Kingdom}
\author{Raymond E. Goldstein}
\affiliation{Department of Applied Mathematics and Theoretical Physics, Centre for Mathematical Sciences, University of Cambridge, 
Cambridge CB3 0WA, United Kingdom}

\date{\today}

\begin{abstract} 
{We report a novel form of convection in suspensions of the bioluminiscent marine bacterium \emph{Photobacterium phosphoreum}. 
Suspensions of these bacteria placed in a chamber open to the air create persistent luminiscent plumes most easily visible when 
observed in the dark. These flows are strikingly similar to the classical bioconvection pattern of aerotactic swimming bacteria, which 
create an unstable stratification 
by swimming upwards to an air-water interface, but they are a puzzle since the strain of \emph{P. phosphoreum} used does not 
express flagella 
and therefore cannot swim.  
Systematic experimentation with suspensions of microspheres reveals that these flow patterns are driven not by the bacteria but by the 
accumulation of salt at the air-water
interface due to evaporation of the culture medium; even at room temperature and humidity, and physiologically relevant salt concentrations, 
the rate of water evaporation is sufficient to drive convection patterns.  A mathematical model is developed to understand the mechanism 
of plume formation, and linear stability analysis as well as numerical simulations were carried out to support the conclusions. While 
evaporation-driven convection has not been discussed extensively in the context of biological systems, these results suggest that the 
phenomenon may be relevant in other systems, 
particularly those using microorganisms of limited motility.
} 

\end{abstract} 

\maketitle

\section{Introduction}
\label{s:intro}

\begin{figure}[h]
\begin{center}
\includegraphics[width=0.6\linewidth]{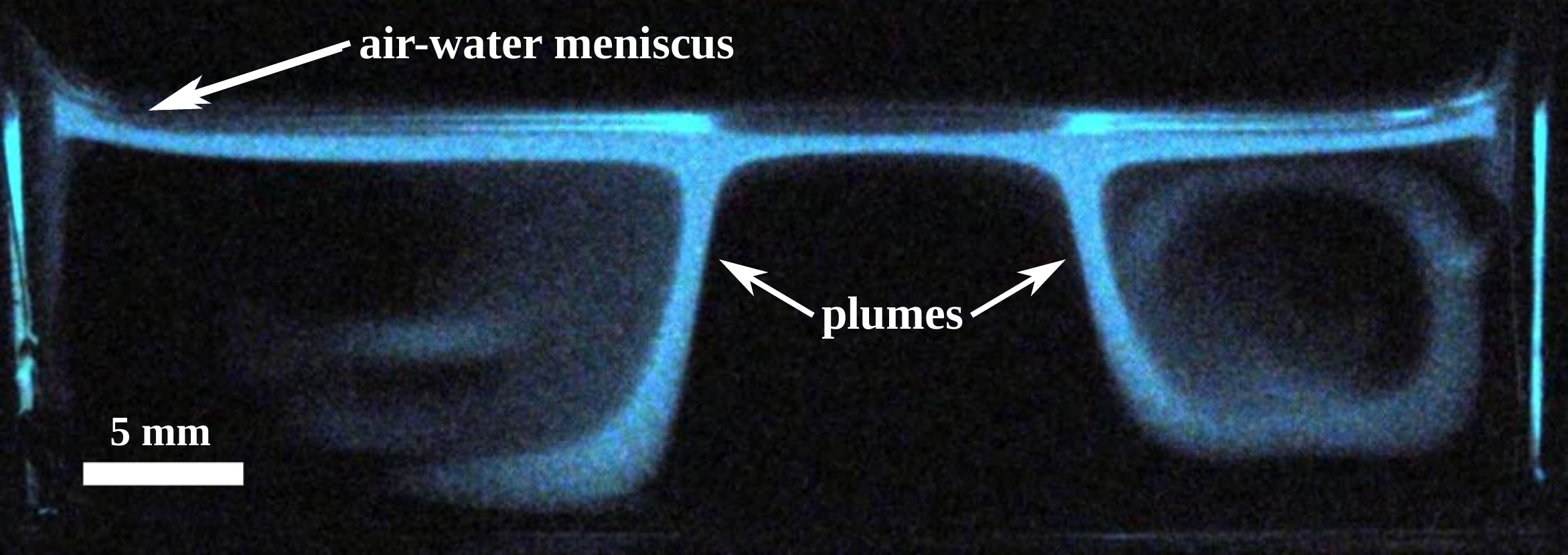}
\caption{Convective flow observed in a suspension of \emph{Photobacterium phosphoreum}. The chamber is $6$ cm long, $1$ mm 
deep and $1$ cm high. Bacteria were cultured to a  density of ~$5.5 \times 10^8$ cells/cm$^{3}$ following the procedure detailed in the 
text. 
This image was acquired with Nikon D300s equipped with a 60 mm f/2.8 macro lens, and an exposure time of 10 seconds.}
\label{plumePPIntro}
\end{center}
\end{figure}

In the deep ocean animals such as certain fish and squid produce light through a symbiotic relationship with bioluminescent 
bacteria \cite{FarmerIII2006}. This luminscence requires oxygen and is regulated by the process of \emph{quorum sensing}, which 
guarantees that 
photons are only emitted when the bacterial concentration is sufficiently high \cite{Dunlap2006}. 
One well-known luminiscent bacterium is \emph{Photobacterium phosphoreum}, a rod-shaped organism $\sim\!3.5$ $\mu$m long 
and $\sim\!0.5$ 
$\mu$m wide, whose bright luminscence can easily be observed when a flask containing
a sufficiently concentrated suspension is swirled gently to oxygenate the fluid.  If instead the suspension is placed in a
cuvette like that shown in Fig. \ref{plumePPIntro} then after several minutes, during which the bacteria deplete the oxygen and 
bulk luminiscence fades away, a novel phenomenon of bright convective plumes is observed.  In the figure, which was taken in the dark, 
regions of blue are luminiscent and have a high concentration of oxygen.  Careful observation shows that the fluid within the plumes
flows downward from the meniscus and the luminscence within gradually fades as the fluid descends and recirculates in convective rolls.

The convection pattern in Fig. \ref{plumePPIntro} appears to be a classic example of bioconvection \cite{Pedley1992}, 
an instability that arises from the 
accumulation of swimming microorganisms at the top surface of a suspension typically in response to chemical gradients or light. As 
most cells are denser that the 
suspending fluid, upwards chemo- or phototaxis creates an unstable nonuniformity in the cell concentration which triggers convective 
motion. Yet, the strain of \pp used in our experiments is non-motile under the given conditions, presumably as a 
consequence of
generations of culturing in rich media in which expression of flagella is not triggered.  Given this, the macroscopic flow in 
Fig. \ref{plumePPIntro} cannot be rationalized as a classical bioconvective instability.

An alternate explanation for the observed convective patterns observed is that a metabolic reaction, for example the chemical reaction 
that produces 
bioluminescence, may release a component that is denser than the background fluid. This idea was motivated by examples of so-called 
{\it chemoconvection} in the methylene-blue glucose system, where the chemical reaction creates a product that alters the local buoyancy 
balance \cite{Pons2000}. Using similar arguments, Benoit \et \cite{Benoit2008} explained the buoyant plumes observed in a suspension of 
non-motile \emph{Escherichia coli} consuming glucose. 

As will be explained in detail below, a series of control experiments allows us to 
discard the hypothesis that a chemical reaction or metabolic activity creates the plumes. Instead, we find that the convection arises 
from evaporation of 
the salty suspension. 
A prior report of evaporation-induced convection was presented by Kang \emph{et al.} \cite{Kang2013}, where flows in a sessile droplet 
of salty water were observed by adding fluorescent beads. They noticed that the flow was first oriented towards the perimeter of the 
drop, in the same way as it moves in the coffee stain phenomenon \cite{Brown1827,Deegan1997}. An accumulation of salt occurrs at the 
surface of the drop, which triggers convective motion.
The proposed mathematical model introduces the ``salinity Rayleigh number" $Ra_s$ by analogy with the Rayleigh number used to 
predict the onset 
convection in a fluid heated from below \cite{Rayleigh1916a}. The salinity Rayleigh number compares the buoyancy generated by the 
accumulation of salt with the stabilizing effects of salt diffusion and fluid viscosity. This concept as been used previously in 
the literature to explain the double-diffusion in the phenomenon of salt fingering \cite{Singh2014}, and it has been also found 
to be essential to understand the accumulation of salt in saline groundwater lakes \cite{Wooding1997}. 

This paper is organized as follows: Sec. \ref{s:exp} describes the experimental setup and quantitative observations of plumes. 
In the next two sections a mathematical model is introduced, which is based on the coupling of the Navier-Stokes equation for the 
evaporating fluid with an advection-diffusion equation for the salt. This model is first studied in Sec. \ref{s:lsa} using linear 
stability analysis, and then numerically using a finite element method in Sec. \ref{s:sim}. A comparison between experiments, linear 
stability analysis and numerical results is presented 
in section \ref{s:comparison}. Finally, conclusions and future work are summarized in Sec. \ref{s:conclusions}.

\section{Experimental metehods and results}
\label{s:exp}

\subsection{Culturing of \pp}

Table \ref{medium} shows the ingredients for the liquid and solid media used in the experiments with \emph{P. phosphoreum}. 
Bacterial cultures were streaked on 
agar plates and kept in an incubator at 19$^\circ$C in the dark, and renewed every two weeks. Liquid cultures were prepared by 
picking a bright colony from an agar 
plate and adding it to $50$ ml of bacterial culture, which was then shaken at $100$ rpm in an incubator in the dark at 19$^\circ$C. 
After $15$ hours of shaking, the 
bacterial suspension reached an optical density OD$_{600nm} \sim 1$. For long term storage, the strain was frozen at $-80^\circ$C 
using the cryoprotective 
medium described by Dunlap and Kita-Tsukamoto \cite{Dunlap2006}, although using 65\% (v/v) glycerol in equal parts worked equally well.
The relation between OD and number of bacteria was found using the technique of serial dilution together with the track method.  
We found that OD$_{600nm}$=1 corresponds to ~$5 \times 10^8$ bact/ml.

\begin{table}[h]
\begin{center}
\caption{Liquid and solid medium used to grow \emph{P. phosphoreum}.}
\begin{tabular}{ |l|l|l|l| }
\hline
 \multicolumn{2}{|c|}{Liquid medium (for 1L)} & \multicolumn{2}{c|}{Solid medium (for 1L)}\\
\hline\hline
Peptone & 10g & Nutrient broth powder & 8 g \\ 
sodium chloride & 30g & sodium chloride & 30 g \\ 
glycerol & 2 g & glycerol & 10 g \\ 
di-potassium hydrogen phosphate & 2 g & calcium carbonate & 5 g \\
magnesium sulphate & 0.25 g & agar & 15 g \\  \hline
\end{tabular}
\label{medium}
\end{center}
\end{table}

\subsection{Setup}

\begin{figure}[h]
\begin{center}
\includegraphics[width=0.6\linewidth]{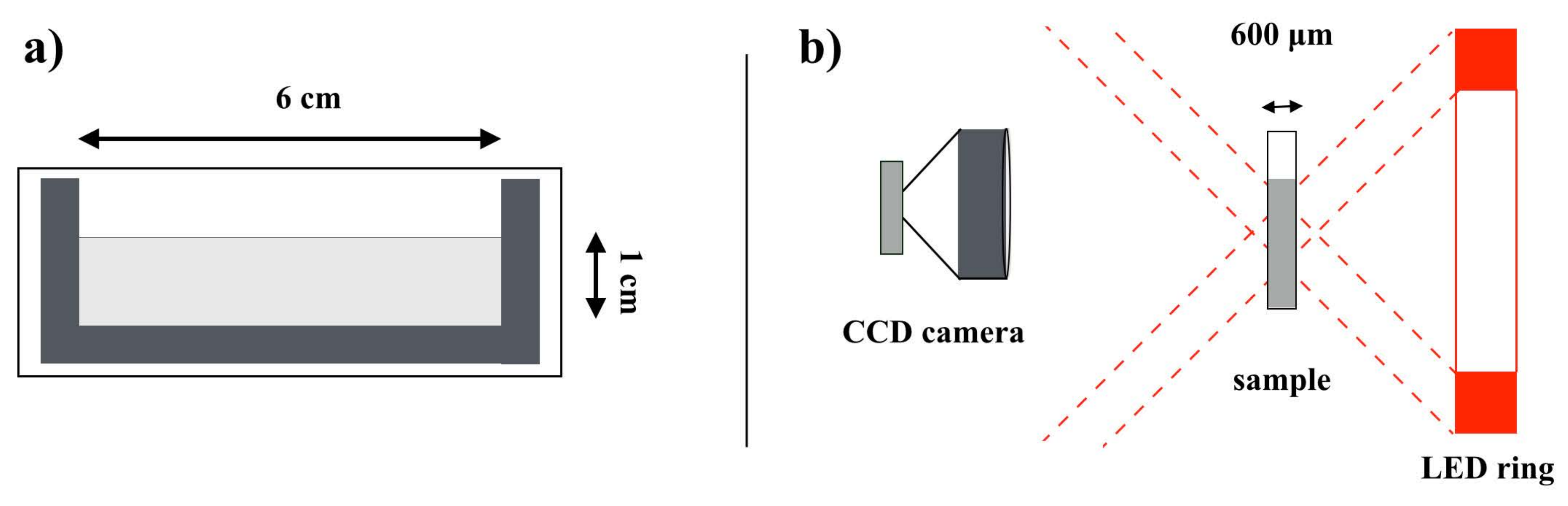}
\caption{Experimental method. (a) Cuvette used in the experiment made with two coverslips held together by 600 $\mu$m thick tape 
cut in a U-shape. (b) Dark field imaging system where the dashed lines represent the light coming from the LED ring.}
\label{expSetup}
\end{center}
\end{figure}

Suspensions were studied in chambers of the form shown in Fig. \ref{expSetup}a, constructed from two glass coverslips (Fisher 12404070) 
held together by two layers of tape 300 $\mu$m thick (Bio-Rad SLF-3001). This tape has internal dimensions $6\times 2$ cm, and one long 
side was cut to create an air-liquid interface. The cuvette, 600 $\mu$m in depth, was filled using a plastic syringe connected to a 
stainless steel 
needle (Sigma-Aldrich CAD4108). The cuvette was filled with sufficient suspension to yield a fluid height of
$1$ cm, which is the initial 
condition for the model presented in Secs. \ref{s:lsa} and \ref{s:sim}.
Figure \ref{expSetup}b is a schematic representation of the dark field setup used to observe the plumes.  From left to right, the 
elements in this setup are a charge-coupled device (CCD) camera (Hamamatsu C7300) to capture dark field images, the sample, and a red 
light-emitting diode (LED) ring (CCS FPR-136-RD). The camera is placed in the dark spot inside of the cone formed by the 
rays coming from the LED ring. Placing a sample where the rays meet will result in light being deflected or scattered by the the 
suspended particles toward the camera. 

\subsection{Bacteria versus beads}

\begin{figure}[h]
\begin{center}
\includegraphics[width=0.70\linewidth]{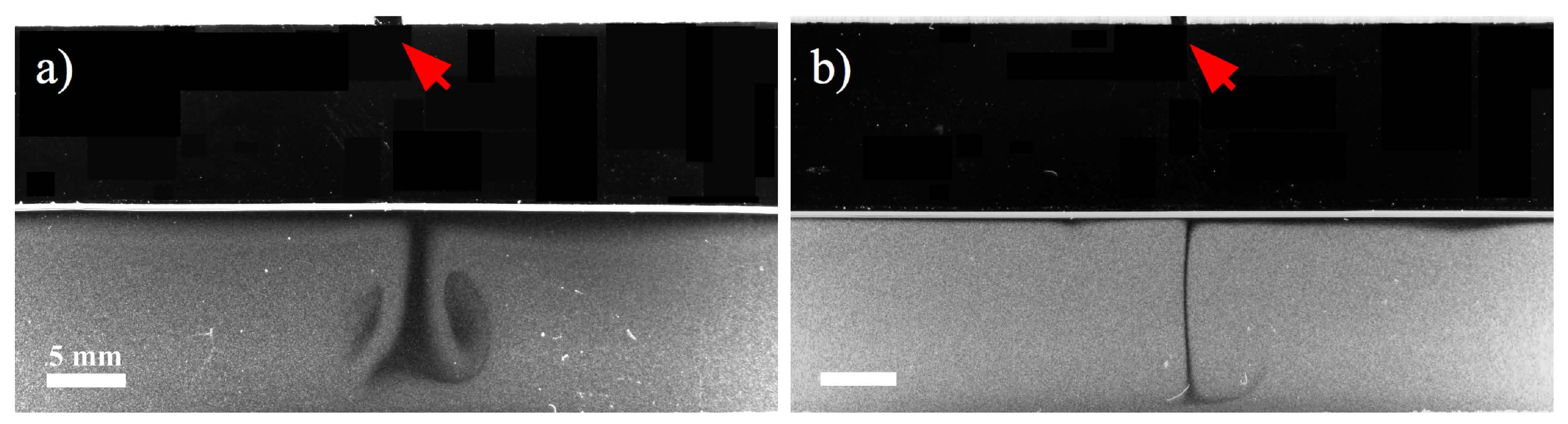}
\caption{Control experiment. \emph{P. phosphoreum} (left) and 3 $\mu$m beads (right) were placed in the cuvettes described in 
Fig. \ref{expSetup} but in this case the top boundary was closed except for a single hole (indicated by red arrows). Both experiments were 
performed at optical density OD$_{600nm}=1.5$ and the images were taken $75$ minutes after the experiment started.}
\label{oneHole}
\end{center}
\end{figure}

The hypothesis that the chemical reaction leading to light production in \pp~ created dense components that trigger hydrodynamic plumes was 
excluded by two control experiments. The first one consisted of placing non-motile, non-bioluminescent bacteria in the same experimental 
setup. A genetically modified strain of \emph{Serratia} (ATCC 39006), provided by Dr. Rita Monson (Department of Genetics, 
University of Cambridge), also created the plumes observed with \pp, even though this bacterium does not produce light 
when it consumes oxygen.
The second and most important control experiment came from using 3 $\mu$m diameter polystyrene beads (Polyscience 18327) at a 
similar optical density 
(OD) as in the experiments with \emph{P. phosphoreum}. Figure \ref{oneHole} shows side by side experiments with 
(a) bacteria and (b) bead suspended {\it in the 
same bacterial medium} and observed using the dark field technique (see Supplemental videos $1$ and $2$ \cite{suppl}).   
In these experiments we sealed the top boundary except for one 
small hole (indicated by an arrow) which allowed evaporation.   We found that the position of the plume is precisely correlated with the position 
of the source of evaporation. Note in particular that the center of the plumes is depleted of bacteria or beads, an important 
feature discussed later.

As the plume generation does not depend on any life processes of the bacteria, in subsequent experiments described below we focused
exclusively on suspensions of microspheres instead of bacteria. Switching to beads also
allowed us to explore a wide range of salt concentrations, far beyond that which is physiologically 
possible using \emph{P. phosphoreum}.

\subsection{Experiments using beads}

\begin{figure}[h!]
\begin{center}
\includegraphics[width=0.7\linewidth]{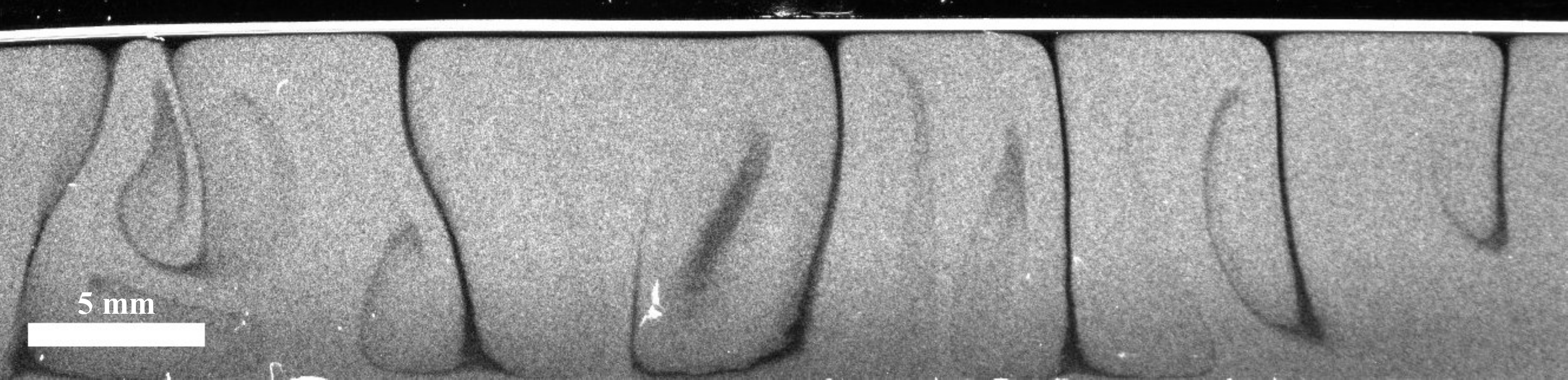}
\caption{Convection in a microsphere suspension. Dark field image of the plumes observed in a suspension of 3 $\mu$m diameter beads 
with 1\% (w/w) salt concentration. The optical density was $OD_{600nm}$=1 and the image was taken 30 minutes after the experiment 
had started.}
\label{1percent}
\end{center}
\end{figure}

Figure \ref{1percent} shows convective patterns observed in a suspension of 3 $\mu$m diameter microspheres in a 1\% (weight/weight) 
salt solution (see also Supplemental video $3$ \cite{suppl}). 
The bright horizontal line at the top of the image is the air-water meniscus. Note that the plumes appear dark in the dark field 
image, which means an absence of particles deflecting light in those regions. This situation is very different from 
conventional bioconvection, in which the plumes appear bright due to the relatively higher bacterial concentration within them.

\begin{figure}[h!]
\begin{center}
\includegraphics[width=0.4\linewidth]{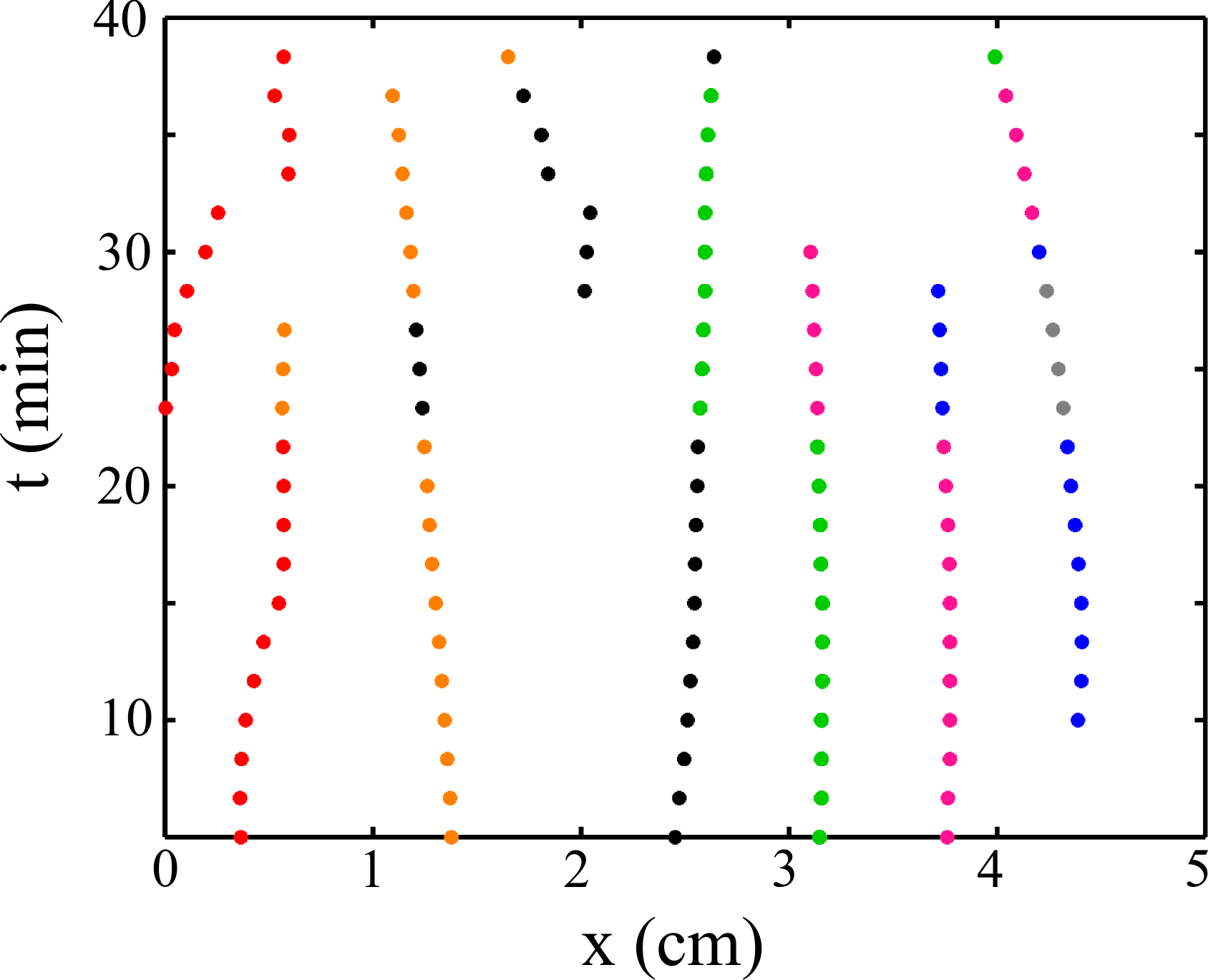}
\caption{Plume dynamics. Graphical representation of the plume position in time for an experiment using 3 $\mu$m diameter 
beads and 1\% (w/w) salt concentration. Different colours indicate the position of the first plume (red), second (orange), 
third (black), fourth (green), fifth (pink), sixth (blue) and seventh (grey).}
\label{representation3m1p}
\end{center}
\end{figure}

A protocol was designed to find the positions of the plumes in time for different salt concentrations.  An image like that in 
Fig. \ref{1percent} was first cropped near the top surface, keeping the horizontal length but narrowing in the vertical direction. 
After converting to black and white with a threshold of $0.5$, the greyscale was inverted. Then, a one-dimensional plot was 
generated by averaging the pixel intensity vertically, where noise was reduced by applying the Matlab function \emph{smooth}. 
The peaks were found using the function \emph{findpeaks} with the condition that the peaks be greater than $0.4$ mm apart. 
Figure \ref{representation3m1p} shows the plume positions corresponding to the conditions of Fig. \ref{1percent}, 
where different colors represent the position of the first plume, second plume, and so on until the seventh plume. In this representation 
a given plume can change color because a new plume appeared at the left side. 

\begin{figure}[h!]
\begin{center}
\includegraphics[width=0.95\linewidth]{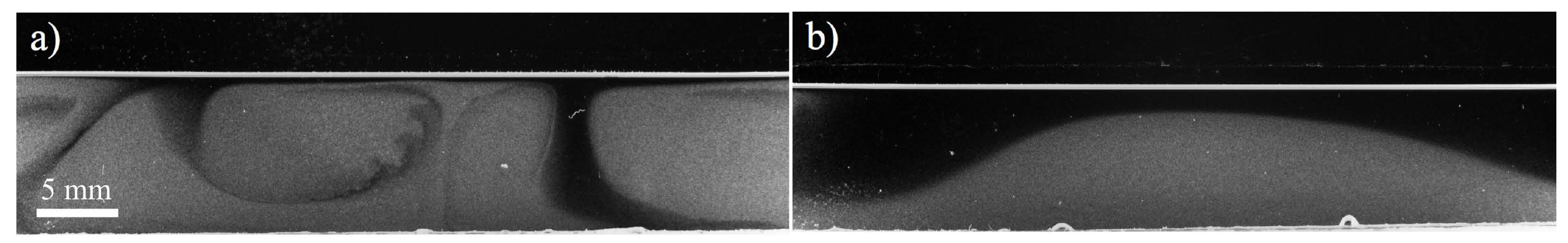}
\caption{Plumes at different salt concentration using $2.17$ $\mu$m polyestyrene beads. (a) $0.1$\% (w/w) salt concentration, $1$ hour after 
beginning of the experiment. (b) $0.01$\% (w/w) salt concentration, $10$ hours after the experiment started. The initial height for both 
experiments was $1$ cm, and in the image on the right the effect of evaporation can be seen by the lower meniscus.}
\label{diffSalt}
\end{center}
\end{figure}

Figure \ref{diffSalt} shows convection in suspensions with (a) $0.1$\% (w/w) salt and (b) $0.01$\% (w/w) salt. Comparing these images to 
that in Fig. \ref{representation3m1p} one can see that for smaller salt concentrations the thickness of the plumes and the spacing 
between them both increase.  When these experiments were analyzed using the code described above, the average distance between 
plumes using 1\% salt concentration was $\lambda_{exp,1\%}=0.67 \pm 0.06 \mbox{ cm}$, while for 0.1\% was 
$\lambda_{exp,0.1\%}=1.58 \pm 0.08 \mbox{ cm}$. 
In 0.01\% suspensions it was difficult to determine the wavelength although it is clearly larger than when using ten times more salt.

Another feature in these experiments is that the less salt in the suspension, the longer it takes to observe the appearance of plumes. 
This can be seen in Fig. \ref{diffSalt} where both images were taken once the instability developed: for 0.1\% it occurred after 1 hour, 
while for 0.01\% a full 10 hours was needed. 

\subsection{Plume mechanism}

\begin{figure}[h!]
\begin{center}
\includegraphics[width=0.6\linewidth]{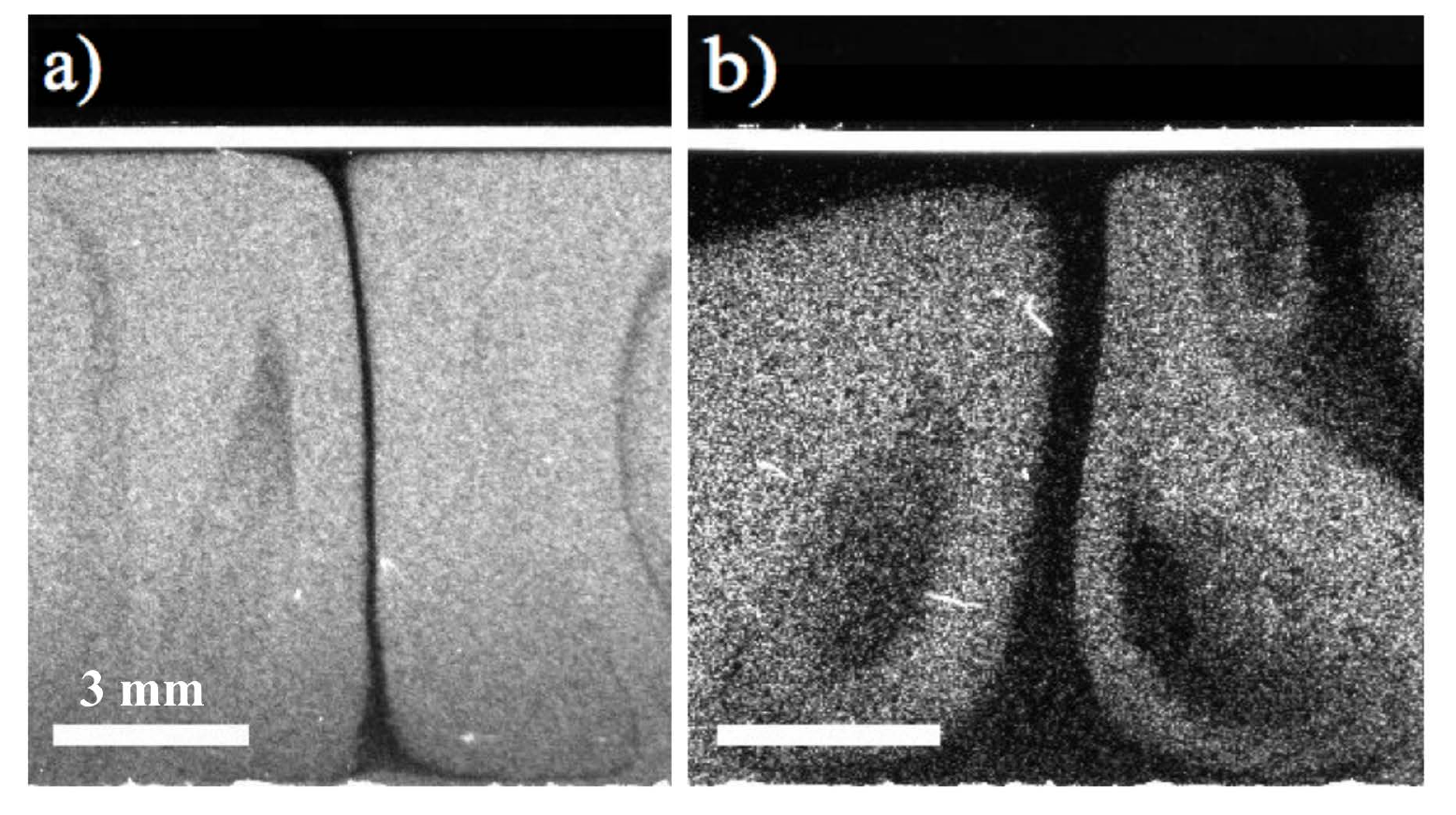}
\caption{Effect of microsphere size.  Plumes using $3$ $\mu$m diameter beads (a) and $6$ $\mu$m beads (b). The material, 
initial height, and salt concentration are the same in both cases. The optical density in both cases is OD$_{600 nm}$=1 and the images were 
taken $45$ minutes after the experiment had started.}
\label{diffSize}
\end{center}
\end{figure}

In the experiments presented above, the plumes appear black in the dark field imaging method, which means that there are no 
beads inside the plumes. 
Our hypothesis is that these dark plumes are a consequence of the sedimentation of beads; while salt is accumulating at the upper
fluid surface due to evaporation the 
beads are continuously sedimenting, and once the instability starts, the convective flows carry fluid down from the top surface, 
which is depleted of beads. 
This idea was investigated by performing the experiments using beads of different sizes, thereby changing the sedimentation speed, while 
keeping the mean salt concentration fixed. Fig. \ref{diffSize} shows the plumes observed using monodispersed polystyrene beads 
$3$ $\mu$m in diameter (Polyscience 09850-5) and $6$ $\mu$m in diameter (Polyscience 07312-5).  It is readily apparent that the 
plumes using larger beads are also wider. 
In this low Reynolds number regime the Stokes law for sedimentation holds, with a speed proportional to (radius)$^2$, so the larger beads
sediment four times faster.   Using ImageJ the thickness of the plumes in both cases were measured at three different points, 
obtaining $\mbox{d}_{3\mu m}=0.152 \pm 0.008 \mbox{ mm}$, while $\mbox{d}_{6\mu m}=0.590 \pm 0.025 \mbox{ mm}$, very 
consistent with the 
ratio of the sedimentation speeds. 

The role of salt in the plume formation was tested by suspending the beads in pure water.  After $27$ hours of observation no plumes 
appeared, confirming that the presence of salt is necessary to create the convection (this experiment was repeated three times). 
From these experiments 
it was also verified that the thermal effect of evaporation is not strong enough to produce plumes under our experimental conditions.
Finally, evaporation was avoided by adding a layer of mineral oil on top of the solution of water and 1\% (w/w) salt, which is the 
same salt concentration used in the experiment in Fig. \ref{1percent}). After $10$ hours no plumes were observed, while the 
air-fluid interface remained fixed. This experiment highlighted the importance of evaporation in accumulating salt at the interface.

In summary, experimental results have shown: a) the center of the plumes is depleted of beads, b) the thickness of the plumes is directly 
related to the sedimentation rate of beads, c) plumes do not appear when beads are suspended in pure water, and d) evaporation is 
needed to observe plumes. Based on these observations, the following mechanism for plume formation is proposed. Initially, the system 
has a nearly homogeneous distribution of beads. Before the instability starts, beads sediment, and an accumulation of salt is built 
up at the air-water interface during this time. When the salt concentration at the top surface has reached a critical value, the 
hydrodynamic instability is triggered by a buoyancy imbalance, carrying down the fluid near the top surface which contains no beads.

\section{Linear stability analysis}
\label{s:lsa}

The hypothesis that accumulation of salt due to evaporation is responsible for plume formation is studied in this section 
using linear stability analysis. A classical example of the use of this technique is in explaining Rayleigh-B\'enard convection, observed
when a fluid is heated from below, in which an instability arises as a consequence of the thermal expansion of the fluid.
B\'enard's observations in 1900 \cite{Benard1900} were followed by Rayleigh's analytical work in 1916 \cite{Rayleigh1916a}, which 
established that the instability takes place when what we now term the Rayleigh number $Ra$ exceeds a critical value.  
$Ra=\beta_T \alpha g d^4/\kappa \nu$ compares the buoyancy force with the dissipative forces (viscosity and thermal diffusion), 
where $\beta_T$ is the temperature gradient, $\alpha$ the coefficient of thermal 
expansion, $g$ the acceleration due to gravity, $d$ the depth of the fluid layer, $\kappa$ the thermal diffusivity 
and $\nu$ the kinematic viscosity. 

The calculation presented here uses a salinity Rayleigh number that compares the buoyancy 
created by salt with the stabilizing effects of salt diffusion and fluid viscosity.  It is 
\beq
Ra_s=\frac{\beta c_0 g d^3}{D \nu},
\eeq
where $\beta$ is the solute expansion coefficient (which gives the change in suspension density due to the salt concentration), 
$c_0$ is the initial salt concentration, and $D$ is the salt diffusion constant. It is important to note that the salinity Rayleigh 
number has often been defined in the literature using the \textit{change} in salinity between top and bottom instead of the initial salt 
concentration (see for example Ref. \cite{Renardy1996}). Yet, even an initially uniform concentration profile will create a gradient 
due to evaporation, so the present definition is more useful here, and was adopted by Kang \emph{et al.} \cite{Kang2013}.
 
The problem analyzed in this section is a laterally infinite two-dimensional cuvette filled with a saline solution. The salt 
concentration is initially homogeneous with a value $c_0$ and the initial height of the column of water is $h_0$ ($1$ cm in 
the experiments). Evaporation has the effect 
of decreasing the fluid height $h(t^*)$ at a constant speed $v_e$, so we assume 
\beq
h(t^*)=h_0-v_e t^*.
\eeq

\subsection{Governing equations} 
\label{eqsChapLinear}

The dynamics of the salt concentration $c^*$ follow the advection-diffusion equation 
\beq
\label{eqSalt}
\parder{c^*}{t^*}+ \nabla \cdot (c^* {\bf u^*})=D {\nabla^2} c^*,
\eeq
with $D$ the salt diffusion constant, and $\vect u^*$ the velocity field.
For the fluid flow, the Navier-Stokes equations with the Bousinesq approximation are used:
\bse
\label{eq1l}
\ba{
\label{eqCdimensions}
\nab \cdot \vect{u^*} &= 0,\\
\label{eqUl}
\parder{\bf{u^*}}{t^*}+\bf{u^*} \cdot \nab \vect{u^*}&
=-\frac{\nab p^*}{\rho_0}+\nu \nab^2 \vect u^*-\frac{\rho}{\rho_0} g \hat{\vect k}.
}
\ese
Eq. (\ref{eqCdimensions}) is the incompressibility condition, and in Eq. (\ref{eqUl}) $p^*$ is the pressure, $\gr_0$ the fluid 
density in the initial state with a homogenous distribution of salt, $\nu$ is the kinematic viscosity, $\gr$ the density of the 
fluid considering the salt distribution, $g$ is the acceleration of gravity, and $\hat {\vect k}$ is the unit vector in the 
vertical direction.

To account for the moving top boundary we normalize the vertical direction by the instantaneous height $h(t^*)$, while the 
horizontal direction is normalized by $h_0$, yielding dimensionless quantities
\ba{
z=\frac{z^*}{h(t^*)}, \hspace{0.5cm} x=\frac{x^*}{h_0}, \hspace{0.5cm} t=\frac{D}{h_0^2} t^*.
\label{dim1}
}
With these definitions and making use of a dimensionless height
\beq 
H=\frac{h(t)}{h_0}, 
\eeq
and the P{\'e}clet number 
\ba{
Pe=\f{h_0 v_e}{D},
\label{dim2}
}
the dynamics of the salt concentration in terms of dimensionless variables is
\beq
\parder{c}{t}=-Pe\frac{z}{H} \parder{c}{z} +\parderd{c}{x}+\frac{1}{H^2} \parderd{c}{z}-u \parder{c}{x} - \frac{w}{H} \parder{c}{z}.
\label{eqCdim}
\eeq
In the equation for the fluid flow, the notation $\vect u=(u,w)$ is introduced to identify the horizontal and vertical components 
of the fluid velocity. The incompressibility condition is
\beq
\parder{u}{x}+\frac{1}{H}\parder{w}{z}=0,
\eeq
and the Navier-Stokes equations for the fluid flow are
\beq
\parder{u}{t}=- Pe \f{z}{H} \parder{u}{z}- u \parder{u}{x}- \f{w}{H} \parder{u}{z}-\parder{p}{x}
+S\!c \left(\parderd{u}{x}+\f{1}{H^2} \parderd{u}{z} \right),
\label{eqUd}
\eeq
in the horizontal direction, and 
\beq
\parder{w}{t}=-Pe \f{z}{H} \parder{w}{z}-u \parder{w}{x}-\f{w}{H} \parder{w}{z}-\f{1}{H}\parder{}{z} \tilde{p}
+S\!c \left(\parderd{w}{x}+\f{1}{H^2} \parderd{w}{z} \right)
- Ra_s S\!c \left(c-1\right),
\label{eqWd}
\eeq
in the vertical direction, where the pressure as $\tilde{p}=p-g z H$ and $S\!c=\nu/D$ is the Schmidt number.
Finally, the rescaled height evolves as
\beq
H(t)=1-Pe\, t.
\label{rescaled_height}
\eeq
Typical values of the experimental and material parameters are given in Table \ref{tableLSA}.

\begin{table}[h!]
\caption{Numerical values of the parameters used in the calculations}
\centering
\begin{tabular}
{c|c|c}\hline 
Parameter & Symbol & Value \\\hline  \hline  
Kinematic viscosity of water& $\nu$ &  $ 10^{-6}$ m$^2$/s\\
Salt diffusion constant & $D$ &  $ 10^{-9}$ m$^2$/s\\
Salt expansion coefficient & $\beta=\rho_0^{-1} \partial{\rho}/\partial{c}$ & 0.007 (w/w)$^{-1}$ \\
Initial height & $h_0$ & 10$^{-2}$ m\\
Evaporation rate & $v_e$ &  $2 \times 10^{-8} $ m/s\\
P{\'e}clet number & $Pe=h_0 v_e/ D$ &  $0.2$ \\
Schmidt number & $S\!c=\nu/ D$ &  $10^3$ \\ \hline
\end{tabular}
\label{tableLSA} 
\end{table}

\subsection{Base state}

Equations (\ref{eqCdim}), (\ref{eqUd}) and (\ref{eqWd}) are now linearized around a base state, which is assumed to
be homogeneous in the $x$ direction:
\beq
\frac{\partial}{\partial x}=0,~~\parder{}{x}\Big{|}_{t=0}=0. 
\eeq
Then, the continuity condition enforces $\partial w/\partial z=0$. Given the boundary condition $w=0$ at $z=0$ and 
$\partial w/\partial z=0$ at $z=1$, the fluid must be at rest, \emph{i.e.} $\mathbf{u}_0(t,z)=(0,0)$. Similarly, 
$c=c_0(z,t)$ satisfies Eq. (\ref{eqCdim}) in the absence of fluid flow,
\beq
\parder{c_0}{t}=\frac{1}{H^2} \parderd{c_0}{z}- \f{P_e z}{H} \parder{c_0}{z}.
\label{eqC0}
\eeq
To find the boundary conditions for $c_0$ it is useful to go back to the dimensionful equations and then translate 
them to dimensionless variables. Integrating the conservation equation along the z-direction,
\beq
\int_0^{h(t)} \left( \parder{c_0^*}{t^*}  - D \f{d^{2} c_0^*}{dz^{*2}} \right) dz^*=0,
\label{continuity}
\eeq
and using the Leibniz integral rule for the first term, we obtain
\beq
\int_0^{h(t)} \parder{c_0^*(z^*,t^*)}{t^*} dz^* = \parder{}{t^*} \left( \int_0^{h(t)} c_0^*(z^*,t^*) dz^*\right)- c_0^*(h(t^*),t^*) 
\, \, \, \frac{dh(t^*)}{dt^*}.
\label{leib}
\eeq
Because the total amount of salt in the system is constant, the first term on the right hand side of Eq. (\ref{leib}) vanishes. 
Then, by using Eq. (\ref{leib}) in (\ref{continuity}), we obtain
\beq
D \frac{d c_0^*}{dz^*}\biggr \rvert_{z^*=h(t)} = - c_0^*(h(t^*),t^*) \, \, \,  \frac{dh(t^*)}{dt^*} =  c_0^* v_e \biggr \rvert_{z^*=h(t)},
\label{leibniz}
\eeq
where the constant evaporation rate $v_e$ was identified. Returning to the dimensionless variables, the upper boundary condition is
\beq
\label{eqFlux}
\frac{d c_0}{dt}\biggr \rvert_{z=1}  = Pe H c_0 \biggr \rvert_{z=1}.
\eeq
Similarly, at $z=0$ the boundary condition of zero flux implies $d c_0/dz=0$. Finally, the initial condition for the salt 
concentration is a homogenous distribution, therefore $c_0(z,0)=1$.

\begin{figure}[h]
\begin{center}
\includegraphics[width=0.4\linewidth]{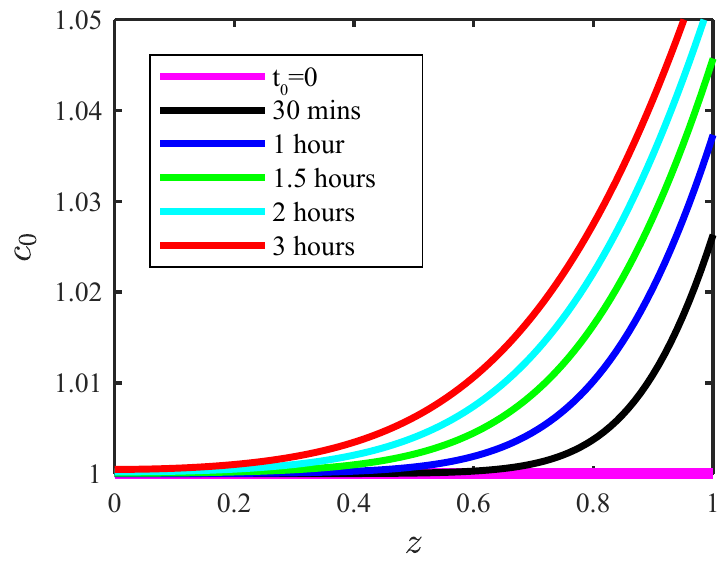}
\caption{Base state.  Normalized salt concentration as a function of the normalized $z$-direction. The plot corresponds 
to the numerical solution of Eq. (\ref{eqC0}) with no-flux boundary conditions (\ref{eqFlux}) and no flux at $z=0$, with $Pe=0.2$, 
calculated from experimental parameters. The colors correspond to different elapsed times since start of evaporation.}
\label{c0Matlab}
\end{center}
\end{figure}

Equation (\ref{eqC0}) with the initial and boundary conditions described above was solved numerically using the function ``pdepe" \cite{pdepe} 
in Matlab, yielding an approximate solution on a given one dimensional grid. 
The dimensionless concentration $c_0$ as a function of the dimensionless vertical direction $z$ is plotted in Fig. \ref{c0Matlab}, 
where different colours represent the salt profiles after a given numbers of hours. In the figure it is possible to observe 
how the salt accumulates near $z=1$, and how the overall concentration increases as a result of the decrease in height while 
the total amount of salt is conserved.    

\subsection{Linearized equations for small perturbations}
\label{secLinear}

The timescale required to evaporate the suspension completely, around 50-100 hours, is much longer than the tens of minutes over which 
plumes develop, and therefore it can be assumed that the timescale $t$ of the plume dynamics is much shorter than that 
needed to reach a base state ($t_0$). The first step to linearize Eqs. (\ref{eqCdim}), (\ref{eqUd}) and (\ref{eqWd}) is 
to consider a small perturbation around the base state at order $\varepsilon$ such that
\bse
\ba{
\vect u(\vect x,t)&=\vect 0 + \varepsilon \vect u' (\vect x, t)+O(\varepsilon^2),\\
c(\vect x,t)&=c_0 (z,t_0) + \varepsilon c' (\vect x, t)+O(\varepsilon^2),\\
p(\vect x,t)&=p_0 (z,t_0) + \varepsilon p' (\vect x, t)+O(\varepsilon^2),
\label{pert}
}
\ese
where $\vect u'=(u', w')$ is the perturbed velocity field, and $c'$ is the perturbed salt concentration. 
Substituting Eqs. (\ref{pert}) into Eqs. (\ref{eqCdim}), (\ref{eqUd}) and (\ref{eqWd}), and collecting the terms at $O(\varepsilon)$, 
we obtain
\bse
\ba{
\label{eqc}
\parder{c'}{t}&=-P_e \f{z}{H} \parder{c'}{z}+\parderd{c'}{x}+\frac{1}{H^2} \parderd{c'}{z}-\frac{w'}{H} \parder{c_0}{z},\\
\label{eqinc}
0&=\parder{u'}{x}+\frac{1}{H}\parder{w'}{z},\\
\label{equ}
\parder{u'}{t}&=-Pe\f{z}{H} \parder{u'}{z}-\parder{p'}{x}+S\!c \left(\parderd{u'}{x}+\f{1}{H^2} \parderd{u'}{z} \right),\\
\label{eqw}
\parder{w'}{t}&=-Pe\f{z}{H} \parder{w'}{z}-\f{1}{H} \parder{p'}{z}+S\!c \left( \parderd{w'}{x}+\f{1}{H^2} \parderd{w'}{z} \right) -S\!c Ra_s ~ c'.
}
\ese

Since the evaporation speed is so small, the variation of height can be neglected in the perturbed equation. Thus, at any given 
moment $t=t_0$, $H(t)=h(t)/h_0$ can be treated as a constant in time such that $H(t)=H(t_0)\equiv H_0$.
Equations (\ref{eqc})-(\ref{eqw}) can be reduced to two coupled equations by introducing the streamfunction $\psi(x,z,t)$, which satisfies 
\beq
\label{stream}
u'=\f{1}{H_0}\frac{\partial \psi}{\partial z}, \hspace{1cm} \mbox{and} \hspace{1cm} w'=-\frac{\partial \psi}{\partial x}.
\eeq
Taking $(1/H_0)\partial/\partial z$ of Eq. (\ref{equ}) and subtracting $\partial/\partial x$ of Eq. (\ref{eqw}), we obtain
\beq
\parder{}{t} \hat{\lap} \psi = Pe \f{z}{H_0} \parder{}{z}\hat \lap \psi + S\!c \hat \lap^2 \psi + Ra S\!c \parder{c'}{x},
\label{eqPsi}
\eeq 
where $\hat \lap=\partial^2/\partial x^2+(1/H_0^2)\partial^2/\partial z^2$. Then, the equation for the perturbed salt concentration 
in terms of $\psi$ becomes
\beq
\parder{c'}{t}=-Pe\frac{z}{H_0} \parder{c'}{z} + \parderd{c'}{x}+\frac{1}{H_0^2}\parderd{c'}{z}+\frac{\partial \psi}{\partial x} \parder{c_0}{z}.
\label{eqC}
\eeq

To find the solution for these coupled equations, a normal-mode solution is considered
\beq
\label{modal}
\psi(x,z,t)=\hat{\psi}(z)e^{(ik x + \sigma t)} + c.c, \hspace{0.5cm}  \mbox{and} \hspace{0.5cm} c'(x,z,t)
=\hat{c}(z)e^{(ikx + \sigma t)} + c.c,
\eeq
where $k$ gives the modulation of the pattern in the horizontal direction and the sign of $\sigma$ indicates the 
stability of the solution. Using this solution, Eqs. (\ref{eqPsi})-(\ref{eqC}) can be written as a linear system
\beq
\label{linearSystem}
\sigma \left(\begin{array}{cc}  D^2-k^2 & 0 \\ 0 & I \end{array}\right) \left(\begin{array}{c}\hat \psi \\ 
\hat c\end{array}\right)=\left(\begin{array}{cc}\mathbb{A} &  ik Ra Sc \\ i k \frac{d c_0}{dz} 
& \mathbb{B}\end{array}\right) \left(\begin{array}{c}\hat \psi \\\hat c\end{array}\right),
\eeq
with $D=(1/H_0)d/dz$, and the operators $\mathbb{A}$ and $\mathbb{B}$ are 
\bse
\ba{
\label{opA}
\mathbb{A}&=Pe z D \left(D^2 - k^2 \right) + S\!c (D^2 - k^2)^2, \\
\label{opB}
\mathbb{B}&=-Pe z D + (D^2 - k^2).
}
\ese

In the linear system, $dc_0/dz$ is the numerical derivative of $c_0(z,t)$, which is found by solving Eq. (\ref{eqC0}). To find the 
boundary conditions 
we recall that the perturbed fluid velocity is $\vect u(\vect x,t)=\vect 0 + \varepsilon \vect u' (\vect x, t)$. At $z=0$, $\vect u=0$, this yields
\beq
u'=\frac{1}{H_0} \f{d \psi}{dz} = 0, \hspace{0.5cm} \mbox{and} \hspace{0.5cm} w'=-i k \psi = 0.
\eeq 
Since $\psi=\hat{\psi}(z) e^{ikx + \sigma t}$, these conditions imply $\hat \psi=d \hat \psi/dz=0$ at $z=0$. At $z=1$ 
there is a free surface 
and no vertical velocity, $w'=0$, which implies $\hat \psi =0$. Similarly, imposing the zero-stress condition,
\beq
\frac{1}{H_0}\parder{u'}{z}+\parder{w'}{x}= 0,
\eeq 
which in terms of the streamfunction $\psi$ is equivalent to $(1/H_0^2) d^2 \psi/dz^2=0$. Thus, at $z=1$, $\hat \psi=d^2 \hat \psi/dz^2=0$. 

\subsection{Numerical implementation}

A numerical solution to the governing equations begins with
the solution of Eq. (\ref{eqC0}), which yields the salt concentration profile $c_0(z)$ at a given time $t_0$. Then, 
Eq. (\ref{linearSystem}) is solved by discretizing the interval $z=[0,1]$ into $N=100$ equally spaced nodes, and the derivatives 
are calculated using second-order central finite differences. Since the matrices have size $2N \times 2N$, there are $2N$ eigenvalues and 
$2N$ eigenvectors, but because the equations for the boundary conditions were explicitly written in the matrices, there are six 
spurious eigenvalues, which can be easily discarded from the final solution. We identify the eigenvalue with the largest real part,
as this corresponds to the fastest growing mode. The corresponding eigenvector contains the information about $\hat{\psi}$ and 
$\hat{c}$ for the mode which leads to the instability. 
In order to test the code, the traditional Rayleigh-B\'enard convection was solved for rigid-rigid boundary conditions, obtaining a 
neutral stability curve that compares very well to the standard result \cite{Drazin1981}.
The physical parameters used to solve the linear system are shown in Table \ref{tableLSA}. The value of $\beta$ was taken from 
Kang \emph{et al.} \cite{Kang2013}, and the evaporation speed was measured experimentally.

\subsection{Results}

\begin{figure}[b]
\begin{center}
\includegraphics[width=0.50\linewidth]{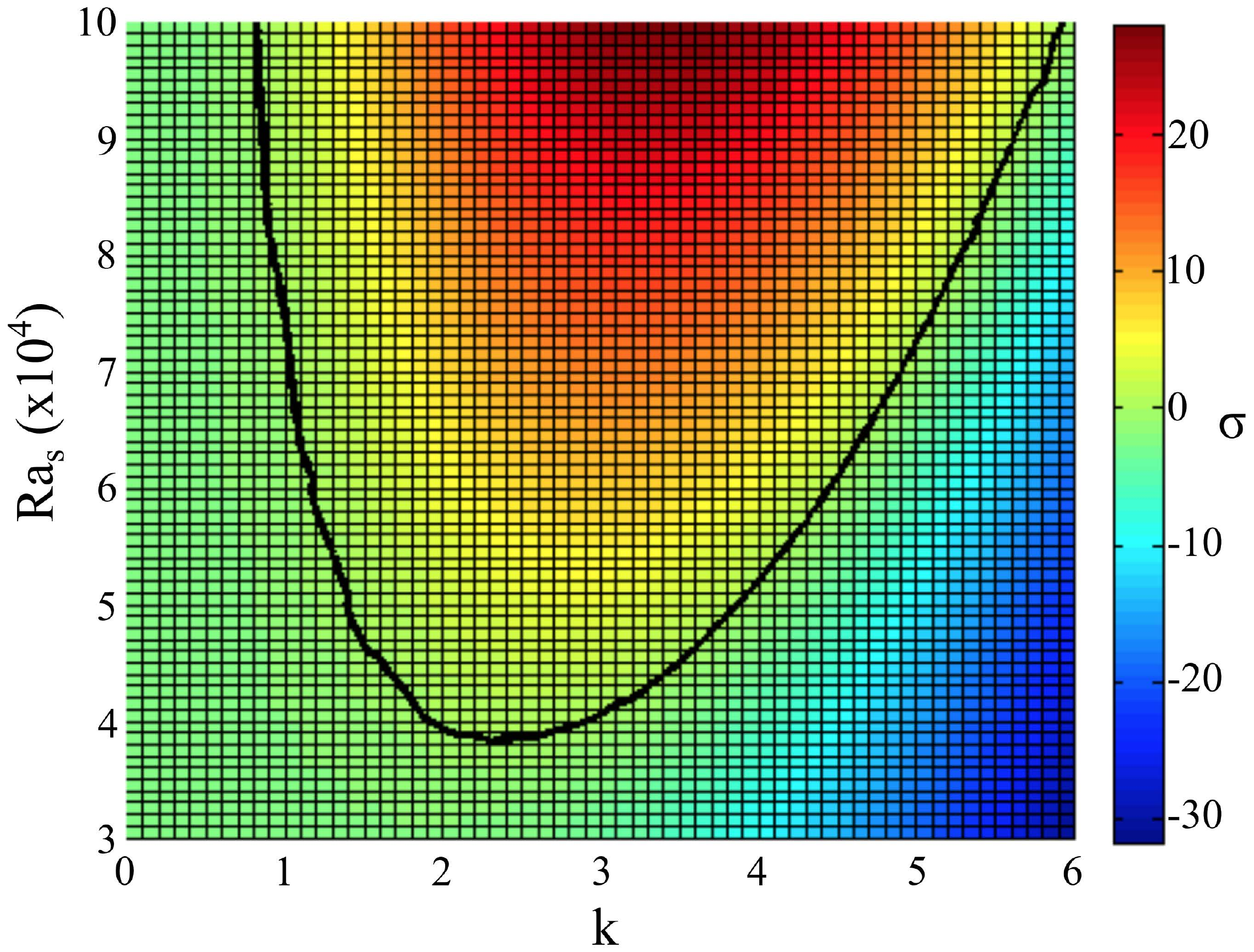}
\caption{Stability analysis.  Values of the largest eigenvalue $\sigma$ in the plane of salinity Rayleigh number $Ra_s$ and wavevector
$k$. In black is the neutral stability curve. Parameters are those in Table \ref{tableLSA}.}
\label{neutral}
\end{center}
\end{figure}

Considering the salt profile at $t_0=$ 1 h, the largest eigenvalues were calculated for different wavevectors $k$ and 
salinity Rayleigh numbers. Figure \ref{neutral} shows the $k-Ra_s$ plane color-coded by the value of the growth
rate $\sigma$, and the neutral stability curve along which $\sigma=0$.
The smallest critical Rayleigh number is $Ra_s^*=3.8 \times 10^{4}$, with a critical wavenumber $k^*=2.3$.
From the definition of the salinity Rayleigh number, this occurs at a salt concentration of $5 \times 10^{-4}$\% (w/w), and 
implies a critical wavelength of $2.73$ cm. A comparison with experiments is given in Section \ref{s:comparison}.

\subsubsection{Varying Schmidt and P{\'e}clet numbers}

\begin{figure}[h]
\begin{center}
\includegraphics[width=0.85\linewidth]{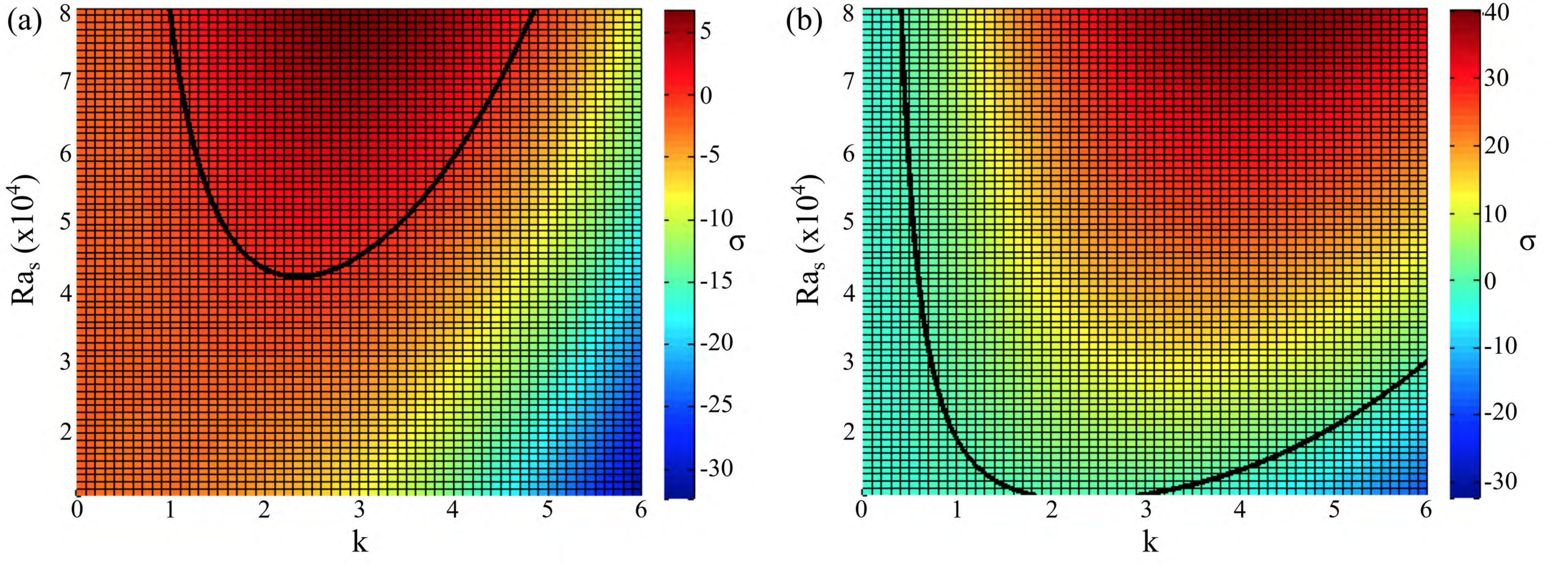}
\caption{Effect of P{\'e}clet number on the neutral stability curve. (a) $Pe=0.1$ and $Sc=1$, (b) $Pe=0.4$ and $Sc=1$.}
\label{Pe}
\end{center}
\end{figure}

The results above were obtained for $Pe=0.2$ and $Sc=1000$. When $Sc=1$, the neutral stability curve does not change with respect 
to the one with $Sc=1000$, but the eigenvalues are different. This result is also obtained in the 
Rayleigh-B\'enard instability, in which case it is possible to demonstrate analytically that the Schmidt number does not 
affect the condition for stability but it does affect the magnitude of the eigenvalues \cite{Drazin1981}. 
Conversely, and as shown in Fig. \ref{Pe}, a higher P{\'e}clet number has the effect of lowering the critical Rayleigh number as well making 
the neutral stability curve broader. These results are expected from the definition of the P{\'e}clet number $Pe=h_0 v_e/D$. A higher 
value can be achieved for example by increasing the evaporation rate, which increases the salt accumulation at the top. 
From the plots in Fig. \ref{Pe} one can see that a higher P{\'e}clet number also has the effect of increasing the value of 
the largest eigenvalue, which results in the instability appearing sooner.

\subsubsection{Varying the time at which the base state is calculated: $t_0$}

All the results shown above were obtained by considering that the base state is reached within one hour, independent of the initial 
salt concentration; the salt profile has been calculated according to Eq. (\ref{eqC0}) and evaluated at time $t_0$ (=1 hour). As 
the dimensionless height was assumed to be constant in the linearized equations $H(t_0)=H_0$, the choice of $t_0$ also affects the value 
of $H_0$. Nevertheless, the evaporation speed is so small that the height decreases only $1$ mm in $10$ hours.

\begin{figure}[h]
\begin{center}
\includegraphics[width=0.95\linewidth]{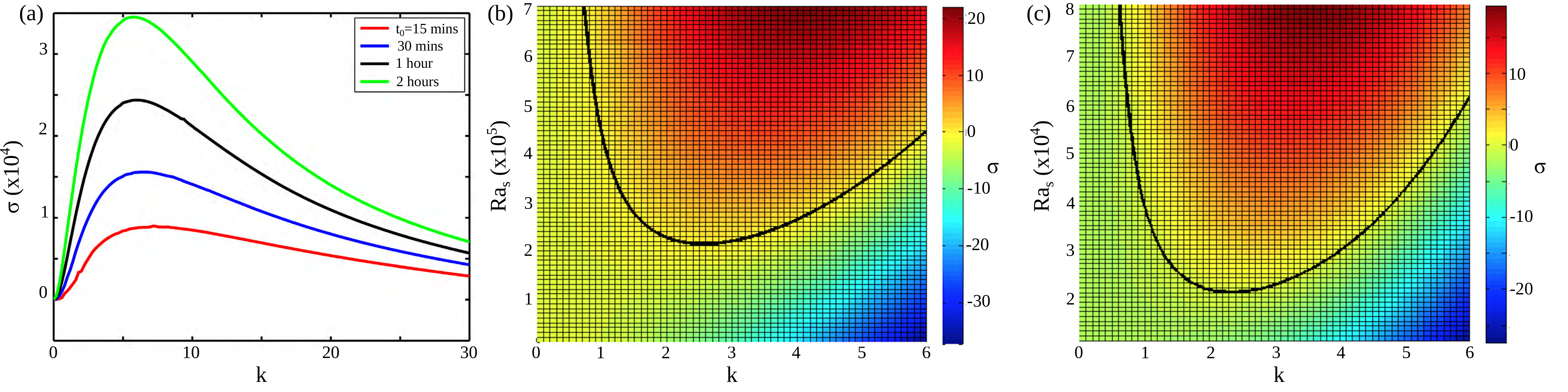}
\caption{Effects of varying $t_0$. (a) Largest eigenvalue for different choices of $t_0$, with $Pe=0.2$ and $Ra_s=6.6 \times 10^7$. 
Neutral stability curves for $t_0=15$ min. (b) and $t_0=2$ hours (c), both with $Sc=1$ and $Pe=0.2$.}
\label{contourWs}
\end{center}
\end{figure}

Figure \ref{contourWs}a shows the largest eigenvalue $\sigma$ as a function of $k$ when the be state is calculated after $15$ min., 
$30$ min., $1$ hour and $2$ hours. As expected, the longer one waits, the more salt has accumulated at the top, and the faster the convection starts.
Nevertheless the position of the peak remains almost the same, so the wavelength of the pattern is not significantly dependent of the choice of $t_0$.
Similarly, the largest eigenvalues changes for different values of $t_0$. Figure \ref{contourWs} shows the neutral stability curve for 
(b) $t_0=15$ minutes and (c) at $t_0=2$ hours.  Comparing these figures to Fig. \ref{neutral} (for $t_0=1$ hour) one observes
that the choice of $t_0$ modifies the value of the critical salinity Rayleigh number without changing the critical wavelength significantly.

\section{Two-dimensional finite element studies}
\label{s:sim}

\begin{figure}[h!]
\begin{center}
\includegraphics[width=0.7\linewidth]{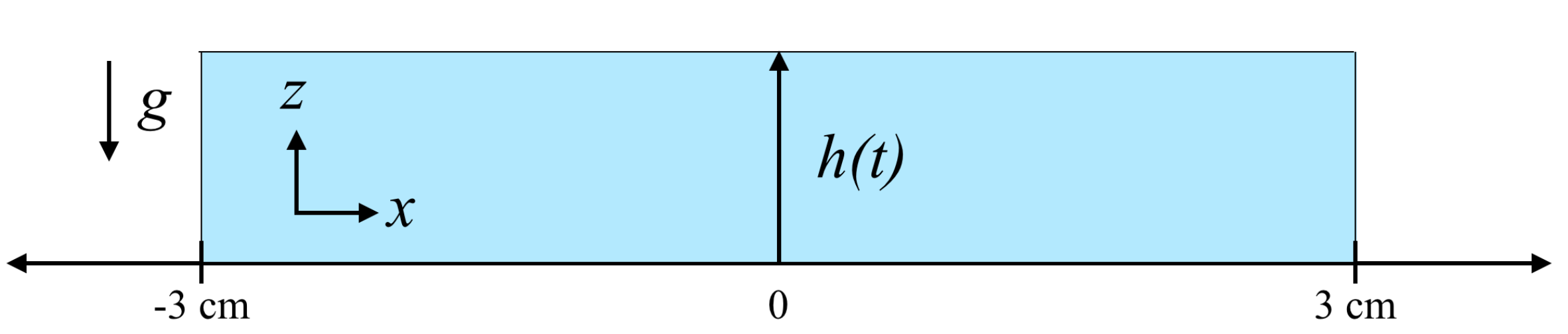}
\caption{Geometry for full numerical studies.  The Navier-Stokes equations are solved in this two-dimensional geometry, 
with dimensions indicated, $g$ the acceleration of gravity and the height $h(t)$ decreasing in time due to evaporation.} 
\label{sketch}
\end{center}
\end{figure}

The formation of plumes was further investigated by performing numerical simulation in the two-dimensional geometry sketched in 
Fig. \ref{sketch}. As in the experiments, the initial height of the column of water is $h_0$= 1 cm, the length of the cuvette is 6 cm, 
and the salt concentration is initially homogeneous and equal to $c_0$.

The numerical studies were performed using the finite element package Comsol Multiphysics.   In this case, the Navier Stokes equation for the fluid 
was coupled to the advection-diffusion equation for the salt concentration, as in previous sections. The program was benchmarked on 
the traditional Rayleigh-B\'enard convection with solid-solid boundaries, obtaining a critical Rayligh number of ~1700, 
in good agreement with the known result of 1707.7 \cite{Drazin1981}.
Comsol can handle the moving top boundary due to evaporation by use of the ``Deformable Geometry" module, which works as follows
: The equations for the fluid flow and salt concentration are written considering a fixed height $h_0$, and an equation for the movement 
of the top boundary is specified. The code then calculates a new mesh by propagating the deformation throughout the domain. Material is added 
or removed depending on the movement of the boundaries, thus the total concentration of species is not conserved between iterations \cite{Comsol}.
As in our case the total amount of salt must be conserved, a flux of salt is added at the top.

For the numerical computations, Equations (\ref{eqSalt}) and (\ref{eqUl})-(\ref{eqCdimensions}) are rescaled using the following 
expressions for length, time, flow speed, pressure and salt concentration:
\ba{
\nonumber
\vect x=\frac{\vect x^*}{h_0}, \hspace{0.5cm} t=\frac{D}{h_0^2}t^*; \hspace{0.5cm} \vect u
=\frac{h_0}{D} \vect u^*,\hspace{0.5cm} p=\frac{h_0^2}{\gr_0 D^2}p^*,\hspace{0.5cm} c=\frac{c^*}{c_0}.
\label{dim1a}
}
The dimensionless equations to be solved numerically are then
\bse
\ba{
\nab \cdot \vect{u} &= 0,\\
\label{dimEqU}
\parder{\vect u}{t}+\vect u \cdot \vect \nabla \vect u &=-\nabla p+Sc  \nabla^{2} \vect u - Ra Sc (c-1) \vect{\hat z},\\
\label{dimEqS}
\parder{c}{t}+\vect u \cdot \nabla c&=\nabla^2 c .
} 
\ese

\subsection{Numerical implementation}

The boundary conditions for the fluid were non-slip at the lateral and bottom sides, and force-free at the top boundary. The buoyancy force 
due to the salt concentration was implemented using the ``Volume Force" feature of Comsol. The boundary conditions for the salt 
concentration were no-flux in the lateral and bottom sides. As stated earlier, when a salty suspension evaporates, water leaves the system, 
but the salt concentration at the top surface increases as this component is conserved in time. This was implemented in Comsol by adding a 
flux of salt at that the top boundary, which was given by $v_e c$, where $v_e$ is the dimensionless rate of evaporation (measured experimentally) 
and $c$ is the dimensionless salt concentration at the top surface. 
The mesh used was the predefine normal mesh calibrated for fluid dynamics, with a maximum element size 0.045 and minimum element size 
0.002. In addition, a corner refinement was added in such way that the elements at the boundaries are scaled by a factor 0.25. The numerical 
values for the parameters in the simulations are the ones shown in Table \ref{tableLSA}.

\begin{figure}[t]
\begin{center}
\includegraphics[width=0.6\linewidth]{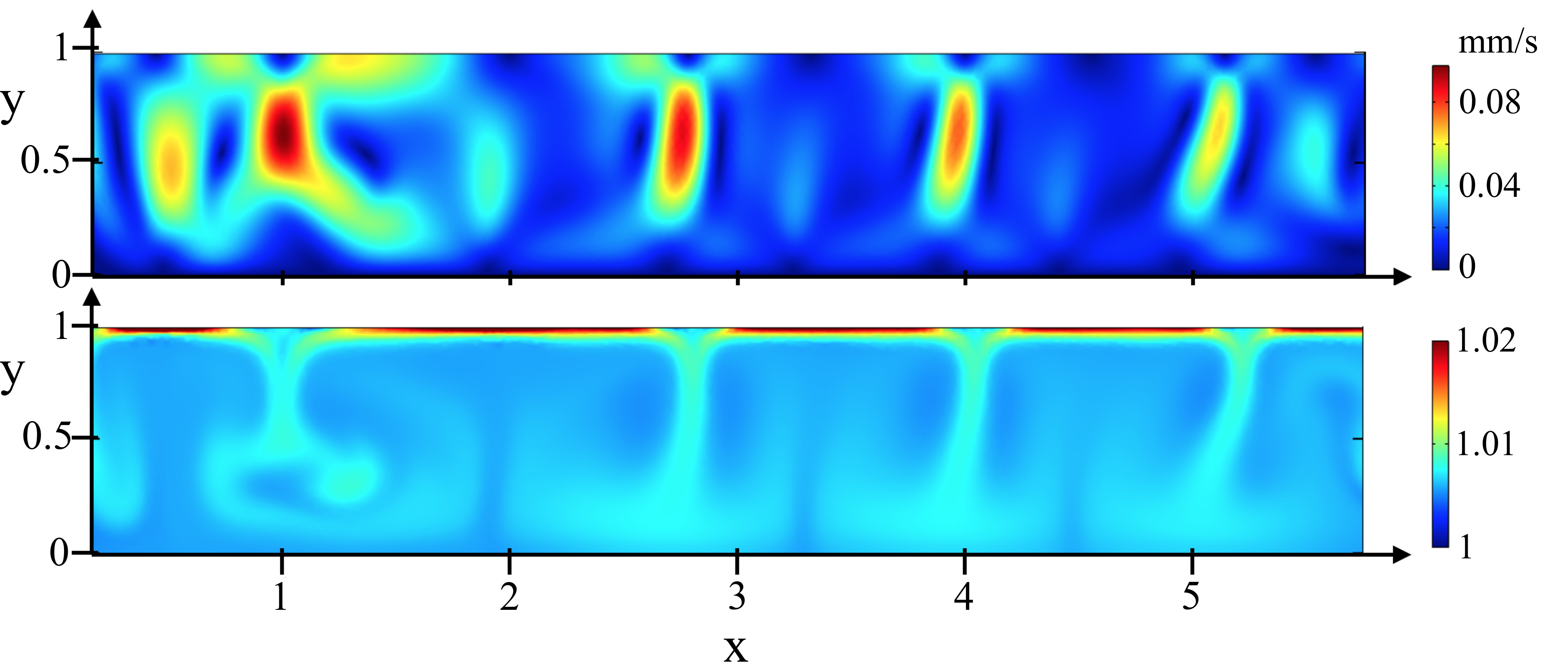}
\caption{Numerical results for 1\% (w/w) initial salt concentration and 60 minutes 
after the experiment has started. (top) flow field; (bottom) normalized salt concentration. }
\label{waterSaltOnly}
\end{center}
\end{figure}

\subsection{Results}

Figure \ref{waterSaltOnly} shows the numerical results for experiments with 1\% (w/w) salt concentration. The flow speed (in mm/s) and the
 salt concentration (normalized by its initial value) are shown 60 minutes after the experiment had started. Here one can see the 
correspondence between high concentrations of salt and the position of the plumes.
These numerics also show the rich plume dynamics noticed in the experiments (see Supplemental videos $4$ and $5$ \cite{suppl}).  
In both cases the number of plumes is not constant and coalescence events are observed. The average distance between plumes can be 
calculated from these images using a similar code to the one described in section \ref{s:exp}, using the color-coding of the velocity field
to identify plumes.  After the image was cropped to narrow it in the vertical direction, it was converted to black and white with a threshold of 
0.2, and the pixels were averaged vertically. This analysis gives a separation between plumes, averaged over time, for 1\% (w/w) salt 
concentration of $\lambda_{sim, 1\%}=0.83 \pm 0.17 \mbox{ cm}$.

\begin{figure}[h]
\begin{center}
\includegraphics[width=0.6\linewidth]{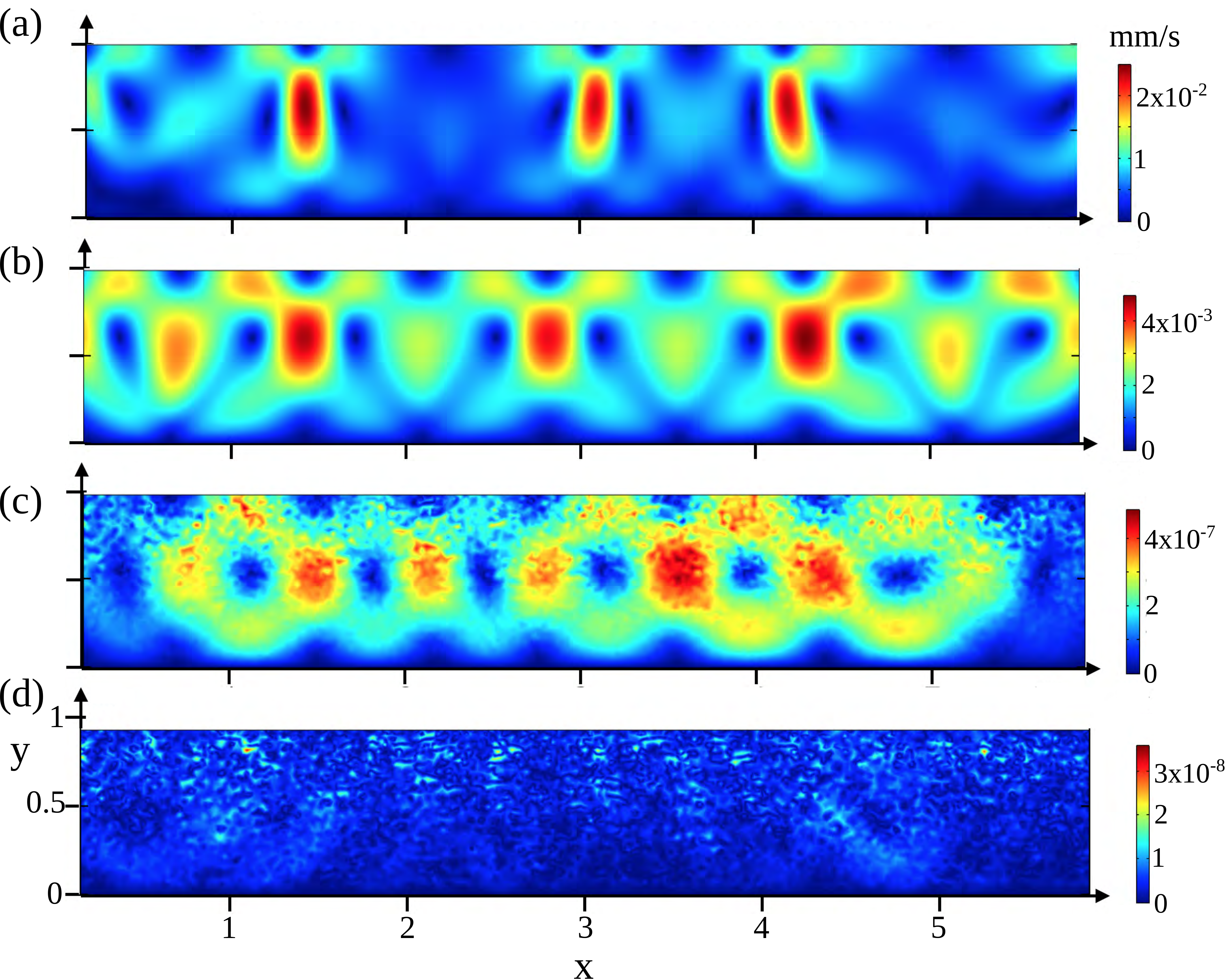}
\caption{Numerical flow fields for different initial salt concentrations. Snapshots at times corresponding to appearance of plumes for
(a) $0.1$\% at $t=45$ min., (b) $0.01$\% at $t=120$ min, (c) $0.001$\% at $t=80$ min, and (d) $0.0001$\% at ten hours, at which point
no plumes were observed.} 
\label{Diffsalt}
\end{center}
\end{figure}

In the experiments it was clear that the lower the overall salt concentration, the longer it takes for plumes to develop, which can be 
understood from the fact that a longer time is needed to accumulate sufficient salt at the top boundary. As can be seen in Fig. \ref{Diffsalt}, 
the same feature is observed in the simulations. Here, the snapshots were taken when the plumes first developed, which is longer as the 
initial salt concentration decreases. In addition, the flow field magnitude also decreases with lower initial salt concentrations. In particular, 
using $0.001$\% (w/w) initial salt concentration, we obtain a typical plume speed of $10^{-4}$ $\mu$m/s, too small to be measured experimentally. 

\section{Comparison between experiments, linear stability analysis and numerical simulations}
\label{s:comparison}

In this section we summarize some of the quantitative measurements of plume obtained using the three approaches.  
We recall that in the linear stability analysis an infinite two dimensional system was considered, while in the experiments and numerical simulations 
the system was laterally finite, so geometric confinement effects can occur.  Moreover, in the experiments, the cuvette of course 
had front and back glass walls, so that the dynamics might be considered more like Hele-Shaw flow that a true two-dimensional system.

\textit{Separation between plumes.}
For 1\% salt concentration experiments, the time-average over several plumes gave a distance between plumes of 
$\lambda_{exp,1\%}=0.67 \pm 0.06 \mbox{ cm}$. From the linear stability analysis, a separation of $\lambda_{linear}=1.05$ cm was 
obtained, and for the same salt concentration, full non-linear numerical simulations gave $\lambda_{sim,1\%}=0.83 \pm 0.17$ cm.
In addition, the linear stability analysis gives information about the distance between plumes at the onset of the instability, which can be compared 
with experimental results before the plumes fully develop. For an initial salt concentration of 0.1\% (w/w), the approximate initial separation 
between plumes is $\lambda_{exp, 0.1\%}=1.58 \pm 0.08$ cm, which compares very well to the value given by the linear stability 
analysis of $\lambda_{linear, 0.1\%}=1.43 \mbox{ cm}$. 

\textit{Minimal salt concentration needed to observe plumes.}
In the experiments, the lowest salt concentration for which we could identify a wavelength was $10^{-2}$\% (w/w). On the other hand, 
the linear stability analysis predicts that the critical salt concentration is $5 \times 10^{-4}$ \% (w/w). This discrepancy can be due to 
multiple factors, including, as mentioned above, the presence of lateral walls and the added viscous drag associated with the
experimental Hele-Shaw geometry. In the numerical simulations, as shown in Fig. \ref{Diffsalt}, no plumes were observed for a salt 
concentration of $10^{-4}$\% (w/w), while for $10^{-3}$\% (w/w), very slow plumes can be identified. Therefore, from this approach 
we obtained that the critical salt concentration is in between $10^{-3}\mbox{ and }10^{-4}$\% (w/w), which is in very good agreement 
with the linear stability analysis result. 

\textit{Time required to observe plumes.}
Experimental observations indicate that when using 1\% (w/w) salt concentration, the instability developed within $15$ minutes, while for 
$0.1$\% (w/w) plumes appeared $2$ hours after the experiment had started. In a conventional 
linear stability analysis, the time needed for the 
instability to start is reflected in the value of the largest eigenvalue, but the in present case
there is also the need to wait a sufficient time for evaporation to create a substantial density
stratification.   Nevertheless, we found that the largest eigenvalue for 1\% (w/w) salt was $7$ times 
larger than for 0.1\% (w/w), remarkably close to the experimental ratio of approximately $8$. 
Similarly, in the numerical simulations 
with $1$\% salt, plumes were observable within $15$ minutes while $45$ minutes were needed to identify plumes in suspensions with $0.1$\% salt.

\section{Conclusions}
\label{s:conclusions}

We have proposed that accumulation of salt due to evaporation can explain the convective patterns observed in suspensions of a 
non-motile marine bacterium. This hypothesis was studied using experimental approaches involving control experiments with 
microspheres of various sizes suspended in growth media with varying salinity.  A mathematical model similar in spirit to that used for 
bioconvection was developed, but focusing on the salt flux created by evaporation at the suspension's upper surface, and 
was studied by linear stability analysis and fully nonlinear numerical simulations. 
The key dimensionless quantity in the model is the salinity Rayleigh number, the critical value of which corresponds to the the threshold of 
salt concentration needed to observe plumes and determines the wavelength at the onset of the instability. Both results compare 
well with experimental observations. Similarly, finite-element method simulations showed a plume dynamics very similar to experimental 
observations, and the dependence on the salt concentration is in agreement with the results obtained using the other two methods. 

The phenomenon of interest here was discovered in a suspension of non-motile marine bacteria, but plumes were also observed using a 
non-marine bacterium (\emph{Serratia}) with low salt concentration in the medium. Therefore the convection does not require salt in 
particular, but merely \emph{some} component that accumulates due to evaporation. The accumulation of solute is always present in 
experiments with an open-to-air surface, and this effect may have been underestimated 
in other systems. For the bacterial medium used, the patterns appeared within $15$ minutes, which may certainly be comparable to the duration of
typical laboratory experiments. On the other hand, when using motile bacteria, the directed movement (chemotaxis) of cells toward the suspension-air 
interface is likely to be much stronger than the convection created by solute accumulation \cite{Tuval2005}. 

In the ocean, an accumulation of salt can generate ``salt fingers", which are formed when cold fresh water surrounds warm salty water.  
This phenomenon is explained by the fact that thermal diffusion is faster than the salt diffusion, which has the consequence that a region 
high in salt will cool before the salt can diffuse, creating a descending plume of dense salty water \cite{Schmitt1994}. In the ocean this 
instability has a profound effect on mixing, which influences factors such as the 
availability of nutrients, heat storage, dispersal of pollutants and the fixation of carbon dioxide, to mention only some effects \cite{Huppert1981}.
An analogy between the convection in the ocean and the fluid flow observed in a suspension of non-motile \pp is particularly interesting, as one 
could argue that the motion created by evaporation may enhance nutrient mixing.  The possibility that bioconvective enhancement of 
mixing may improve population-level growth has been considered previously and found not to occur \cite{Janosi2002}.  The relevance 
of a mixing effect on microorganisms associated with evaporation remains to be studied quantitatively, as does its effect on the so-called 
``sea surface microlayer", the submillimeter interfacial layer within which much important
geobiochemistry occurs \cite{ssmicrolayer}.

\section*{Acknowledgments}

We are grateful to Fran\c{c}ois Peaudecerf for assistance in constructing the bacterial growth curve, Rita Monson for assistance 
with \emph{Serratia} and Andrew 
Berridge for comments on the manuscript. This research was supported in part by ERC Advanced Investigator Grant 247333. 
JD received a PhD fellowship from the Chilean government (Becas Chile). 
KJL was supported by the Basic Science Research Program through the National Research Foundation of Korea (NRF) funded by the 
Ministry of Education (grant number 2016R1D1A1B03930591).

\end{document}